\documentclass[conference]{IEEEtran}
\IEEEoverridecommandlockouts
\usepackage{graphicx}
\usepackage{amsmath}
\usepackage{xcolor}
\usepackage{pdflscape}
\providecommand{\keywords}[1]{\textbf{\textit{Index terms---}} #1}
\usepackage[utf8]{inputenc}
\usepackage[T1]{fontenc}
\usepackage{graphicx}
\graphicspath{ {./images/} }
\usepackage{amsmath}
\usepackage{xcolor}
\usepackage{times}
\usepackage[a4paper, total={6in, 8in}]{geometry}

\usepackage{authblk}

\author[1]{Łukasz Pluszyński (I,III)}
\affil{Cracow University of Technology, Department of Computer Science (I)}
\affil{Cracow University of Technology, Department of Automatic Control and Computer Engineering (II)}
\affil{Quantum Hardware Systems, Lodz (III)}
\author[2]{Krzystof Pomorski(II,III)}

\begin{document}
\onecolumn
\title{Towards construction of analog solver of Schroedinger and Ginzburg-Landau equation based on Long Line}

\maketitle
\begin{abstract}
The analog electronic computers are a type of circuitry used to calculate specific problems using the physical relationships between the voltages and currents following classical laws of physics. One specific class of these circuits are computers based on the interactions between passive circuit elements. Models presented by Gabriel Kron in 1945 are the example of using such passive elements to construct a solver for the problem of free quantum particles confined by rectangular potential. Numerical validation of Kron’s second model is conducted for different shapes of particle confining potential. Kron's model is generalized by introduction of non-linear resistive elements what implies deformation of Schroedinger equation solution into Ginzburg-Landau form.

\end{abstract}

\newpage

\section{Motivation behind development of analog computers modeling quantum systems}
The analog computers are the class of the computers that operate on the continuous signals instead of the discrete numbers such as the most commonly used binary based digital electronics. Their greatest strength of analog paradigm is their ability to utilise the known physical relationships between their components to solve the computationally difficult equations as integro-differential equations without need of reliance on technically costly approximation of real number by discrete states or technical cost of differentiation or integration operators . One of the most commonly known example is the case of use of the relationship between input and output voltages of the operational amplifier with the correctly placed capacitor to get the integrals or the derivatives of the electric signal's function.
\newline \newline
There is inherent ability of the analog electronics to represent various differential equations (and thus referring to wide class of problems) and this potential is being underutilised due to the difficulty of integrating them with the digital systems, analog electronics susceptibility to the interference and the constantly present question of the occurrence of effectively built-in parasitic resistances. 

Presented work leads to the creation of the netlist model of the analog solver for the Schrodinger equation based on Kron’s second model \cite{Kron} and brings its generalisation via numerical validation for classes of effective potential going beyond rectangular potential.






\section{Kron's second model of the analog hardware solver for Schroedinger Equation}
\begin{figure}[h!]
\includegraphics[width=10cm]{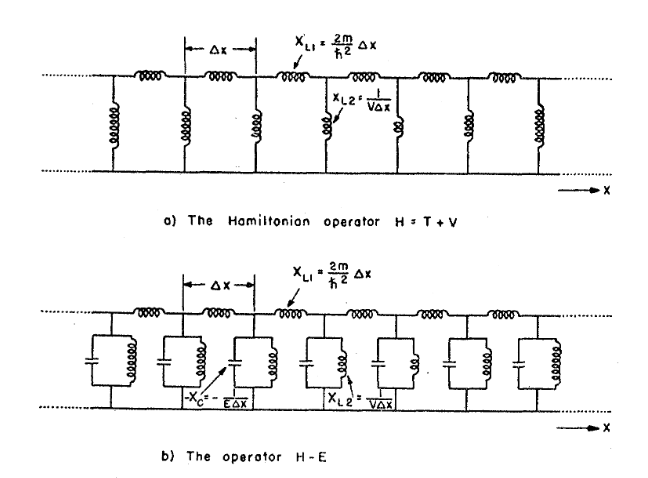}
\centering
\caption{Hamiltonian operator (a) and $H-E$ operator(b) as described by Kron \cite{Kron}.}
\includegraphics[width=10cm]{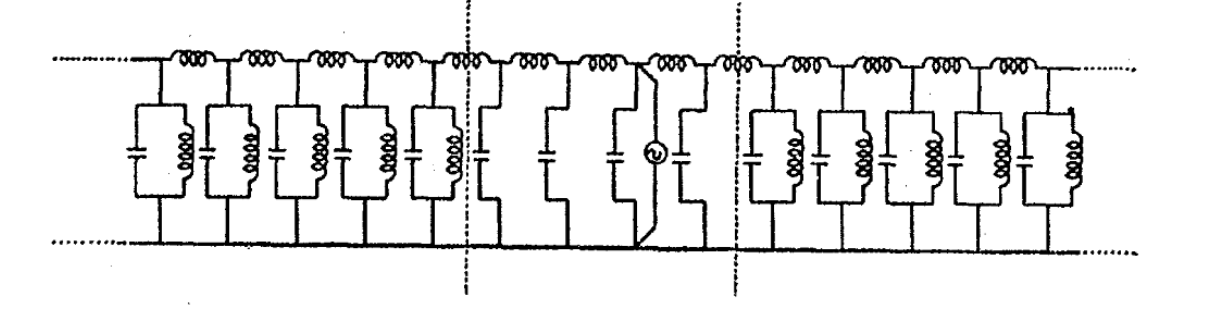}
\centering
\caption{Conceptual schematic of circuit simulating particle in potential well in one dimension with voltage across lower and upper branch as wave-function values as described by Kron \cite{Kron}.}
\label{KronScheme}
\end{figure}

As established by Garbriel Kron in his studies \cite{Kron} we are able to represent the distribution of one dimensional wave-function of Schroedinger equation for single particle in effective potential by classical non-uniform long line with usage of inductance and capacities elements only.
Indeed the second Kron's model points out classical long line with the driving signal (presence of voltage source or current source) of constant frequency and amplitude. In this model we have presence of inductance in horizontal direction, while we set capacitance and inductance in parallel configuration in vertical direction as depicted in Fig.\ref{KronScheme}.
Kron's model leads to representation uses simplistic Hamiltonian of Schroedinger equation of $(\hat{H}(x)-E)\psi(x)=0$ that can be represnted in discrete form as  
\begin{eqnarray}
\Delta x \hat{H}(x) \psi(x) - \Delta x E(t) \psi(x)=0, \hat{H}(x) = \hat{T}(x) + \hat{V}(x), \nonumber \\
\frac{-\hbar^2}{2m }(\psi'(x+\Delta x)-\psi'(x))=\Delta x (E -V_p(x))\psi(x), \nonumber \\
I(x+\Delta x,t)-I(x,t)=\Delta I(x,t)=\frac{V(x)}{Z(x)_{C_1 || L_1}}=(\sqrt{-1}C_1 \omega+\frac{1}{\sqrt{-1} \omega L_1(x)})V(x)= \Delta x (E -V_p(x)), \nonumber \\
\end{eqnarray}
with central preassumption that kinetic energy term in Schroedinger equation is corresponding to electric current phasor change
to $I(x+\Delta x,t)-I(x,t)$ that leads to current phasor $\Delta I(x)$ flowing vertically every lattice step $\Delta x$ in the form
\begin{eqnarray}
\frac{-\hbar^2}{2m}(\psi'(x+\Delta x)-\psi'(x))=I(x+\Delta x,t)-I(x,t)=\Delta I(x,t), \nonumber \\
\end{eqnarray}
what implies
\begin{eqnarray}
(I(x+\Delta x,t)-I(x,t))=\Delta I(x,t)=\frac{V(x)}{Z(x)_{C_1 || L_1}}=(\sqrt{-1}C_1 \omega+\frac{1}{ \sqrt{-1}\omega L_1(x)})V(x)=\Delta x (E -V_p(x)), \nonumber \\
\end{eqnarray}
and under assumption $\psi(x)=V(x)$ we arrive to dependence
\begin{eqnarray}
\frac{-\hbar^2}{2m }(V(x+\Delta x)-V(x))=\Delta I(x), (\frac{d}{dx}V)=-\frac{2m}{\hbar^2} \frac{\Delta I(x)}{\Delta x}=\frac{\Delta Z_L(x)}{\Delta x} I(x)=\sqrt{-1}L \omega  I(x),
\end{eqnarray}
where: $\hat{V}_p$ is potential energy operator,  $\hat{T}$ is  kinetic energy operator equal to $-\frac{\hbar^2}{2m}\frac{d^2}{dx^2}$ or $\frac{p^2}{2m}$,
$\hat{H}$ is Hamiltonian equivalent to  $\hat{V}+\hat{T}$,  E is energy eigevalue,  $\hat{p}$ is momentum operator equivalent to $-i\hbar \frac{ d}{dx}$ and $\Delta x$ represents discrete step in position coordinates, while I(x) and V(x) are phasors of electric current and voltage. We have made explicit assumption that phasor of voltage across upper
and lower branch is representing wave-function. Furthermore in case of long line mimicking Schroedinger equation as by \cite{Dirac}, \cite{QM}, \cite{QM2} we shall imply that $\frac{1}{ \omega L_1(x)}=V_{p}(x)\Delta x$ and that $-C_1 \omega \Delta x =E$ . Under assumption of applicability of impedance concept it turns out that horizontal inductance impedance
$X_L=\frac{2m}{\hbar^2}\Delta x$ is related to kinetic energy operator, vertical inductance  $X_{L1}=\frac{1}{V}\Delta x$ is inverse of potential energy and that impedance of capacitor $-X_C=-\frac{1}{E}\Delta x$ stands for the total energy.
Such results are obtained by derivation of Kron's model conducted in Section IV.
The basic representation of the energy operators in one dimension for time independent Schroedinger equation in Kron's second model is using the serial coils in horizontal direction to represent kinetic energy operator, perpendicular to them inductive coils for representation of potential energy operator and  capacitors connected in parallel to already mentioned vertical inductors for representation of total energy operator E as depicted in Fig.\ref{KronScheme}.


It is worth noting that Figure 2 depicts Kron's representations of Hamiltonian operator (a) and $H-E$ operator (b) for the particle moving in one dimension represented by classical electric circuit.

For purposes of the simulation reflected in analog hardware configuration, the location of the voltage signal generator will be the point of lowest potential energy and in this case represented either by the node with the highest parallel inductance or the node without parallel coil. This setup for basic simulation of particle in rectangular well of potential is depicted in Fig.\ref{CircuitRepresentation}.


The wave-functions $\psi$ being Hamiltonian eigenstate we are looking for can only be electrically measured in invasive way as potential difference between the nodes of upper and lower branch of long line. Due to the relationship between electronic components, the circuit solves the equation only when the current running through the function generator driving the voltage signal equals zero \cite{Kron}.

It is worth noting that we can transit from Schroedinger equation to Ginzburg-Landau equation by
\begin{eqnarray}
\frac{-\hbar^2}{2m}\frac{d^2}{dx^2}\psi(x)+V_p(x)\psi(x)=E \psi(x), \nonumber \\
\frac{-\hbar^2}{2m}\frac{d^2}{dx^2}\psi(x)+\alpha(x) \psi(x)+ \beta(x)|\psi(x)|^2\psi(x)=0, \nonumber \\
\end{eqnarray}
By preassuming transition $V_p(x) \rightarrow (\alpha(x)+ \beta(x)|\psi(x)|^2\psi(x)-E)$ we obtain transition from Schroedinger to Ginzburg-Landau equation. Alternatively we can preassume transition $E \rightarrow V(x)-\alpha(x)-\beta(x)|\psi(x)|^2$ as implementing transition from Schroedinger to non-linear Schroedinger equation. Both transitions can be encoded in certain dependence of $L_1$ and $C_1$ linear density with certain space dependence and with certain voltage dependence. Transition from Schroedinger to Ginzbug-Landau formalims can mean transition from semiconducor to superconductor. Good candidate for circuit elements implementing given transition (Schrodinger to GL) are symmetrized diodes (two diodes connected in antiparallel way) as way of inducing non-linearity to the circuit. Furthermore one can introduce 3 non-linear elements in vertical direction: resistance, capacitance and inductance with x coordinate dependence and one can expect various forms of Ginzburg-Landau like equation solution to emerge in dependence on nature of those elements.
\section{LTspice Simulation of Kron's second circuit}
It is quite straightforward to implement previously defined Kron's model into schematics as depicted in Fig.\ref{CircuitRepresentation}, which was developed and tested in LTspice.
\begin{figure}[h!]
\includegraphics[width=15cm]{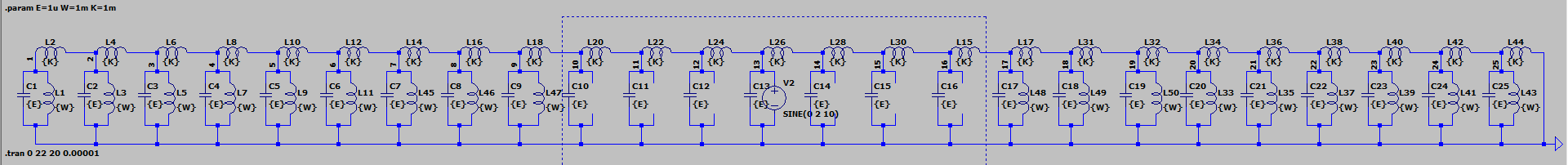}
\centering
\caption{Schematic of twenty five (25) node circuit for simulating particle in rectangular potential well \cite{Dirac}.}
\label{CircuitRepresentation}
\end{figure}
The parameters of the one among many existing optimal simulations found by various trials can be given as follows:
\begin{itemize}
    \item Inductance simulating well of potential:$1mH$
    \item Inductance simulating the kinetic energy operator: $1mH$
    \item Capacitance simulating the total energy:$1uF$
    \item Frequency of the driving signal:$10Hz$
    \item Amplitude of the driving signal:$2V$
    \item Nodes 1 to 25 are points of measurement of voltage and thus level of discretization of continuous position space
    \item Maximal timestep of simulation:$0.00001seconds$
\end{itemize}

The data gathered from simulation took form of the voltage over time in each of tested nodes alongside the current in the node containing the signal generator. 


It should be noted that some deformations of wave-function distribution is observed in simulation results due to comparatively low sampling resolution of the circuit.

\begin{figure}[h!]
\includegraphics[scale=0.3]{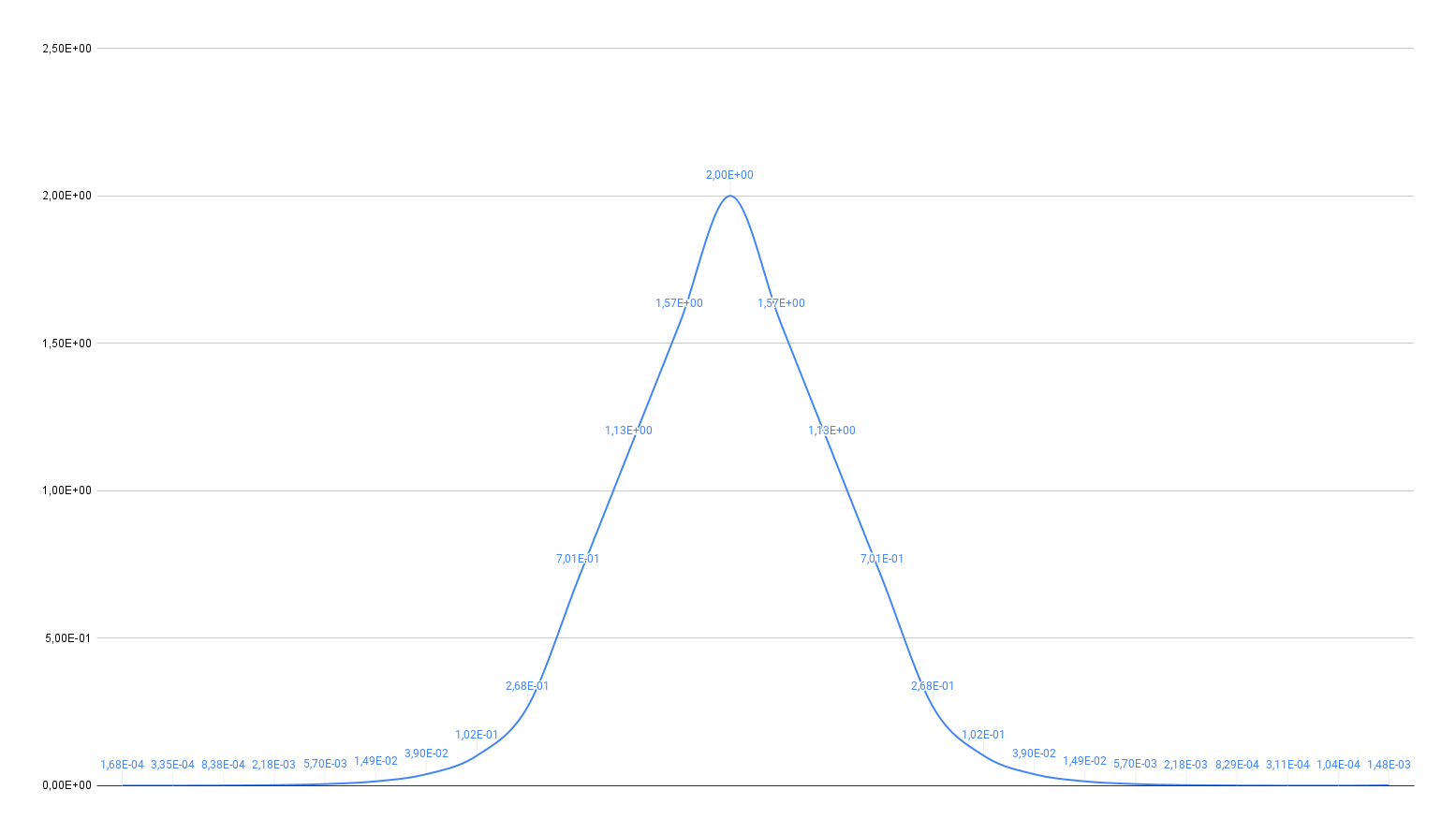}
\centering
\includegraphics[scale=0.3]{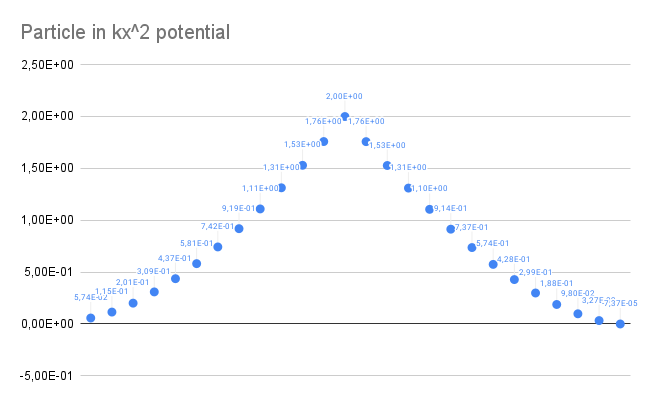}
\includegraphics[scale=0.3]{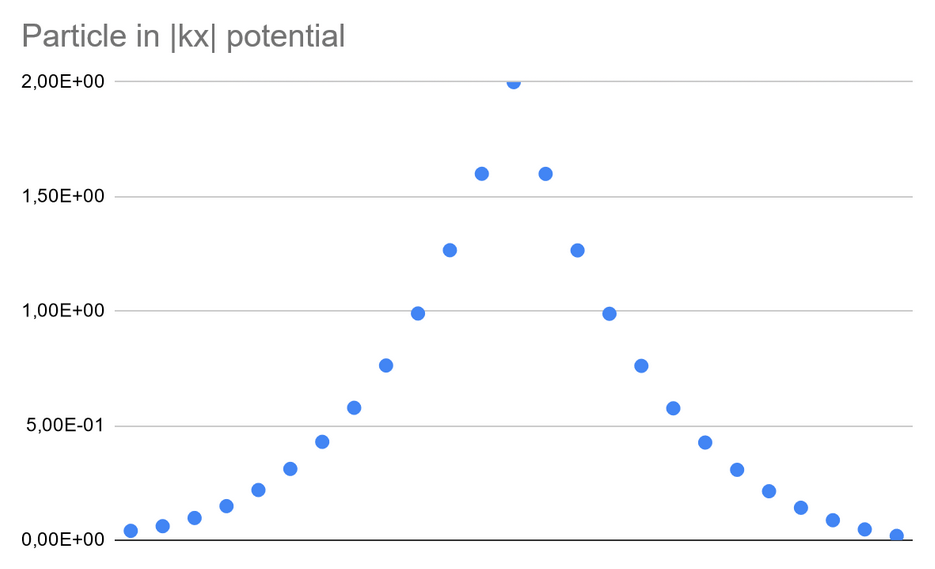}
\label{CircuitRepresentation}
\caption{Obtained numerical simulation results for the particle trapped in rectangular potential well (upper picture), harmonic potential (lower left) and V shape potential (lower right) with use of Kron's second model implemented in LTSpice environment \cite{LTSpice},\cite{AnalogDevices}.}
\end{figure}
Observations about results of simulation experiments can be concluded by following :
\begin{itemize}
    \item Circuit typically needs few cycles to stabilise. This time seems proportional to capacitance's and inductance's of used components.
    \item There is visible spike of voltage at the point where the signal generator is located.
    \item Achieving measurement at the point were current in node V2 equals exactly zero was effectively impossible due to limits of simulation software. This factor can be blamed on final plot being deformed via voltage spike in the node containing the signal generator.
    \item Resultant solution was acceptably close to the expected one quality wise to make a decision about conducting the further experiments with use of stronger hardware for simulation.
\end{itemize}

\section{Derivation of Kron's model expressed by long-line model and its generalization towards Ginzburg-Landau equation}
Non-dissipative Schroedinger equation having real value potential and eigenenrgy states can be expressed by Kron's model with inductance in horizontal direction as well as by capacitance and non-uniform inductances in vertical direction.

\begin{figure}
\centering
\includegraphics[scale=0.7]{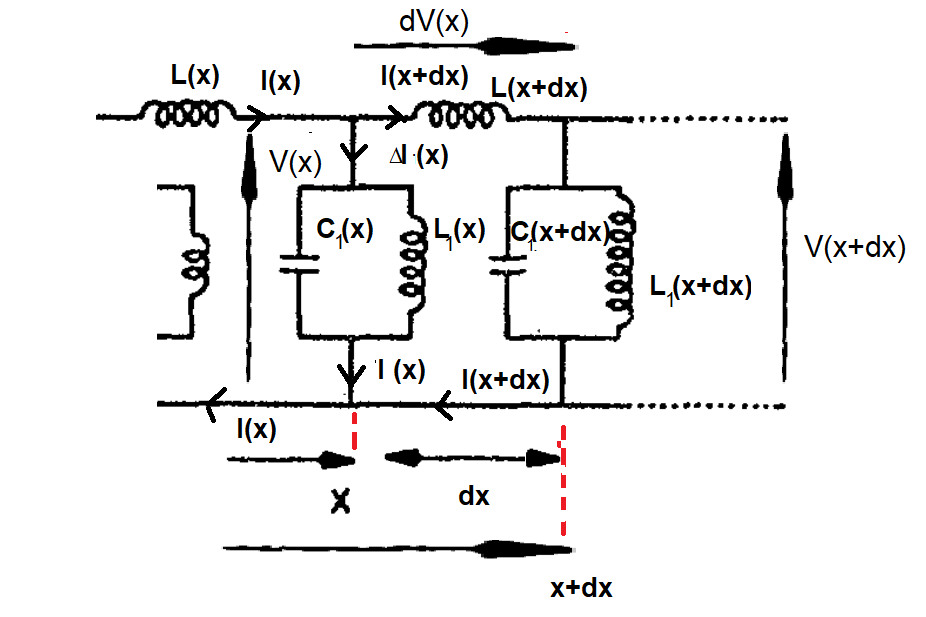}
\caption{Krons model in relation to non-dissisipative non-uniform long line model based on passive linear elements that are space dependent: inductance and capacitance.}
\includegraphics[scale=0.7]{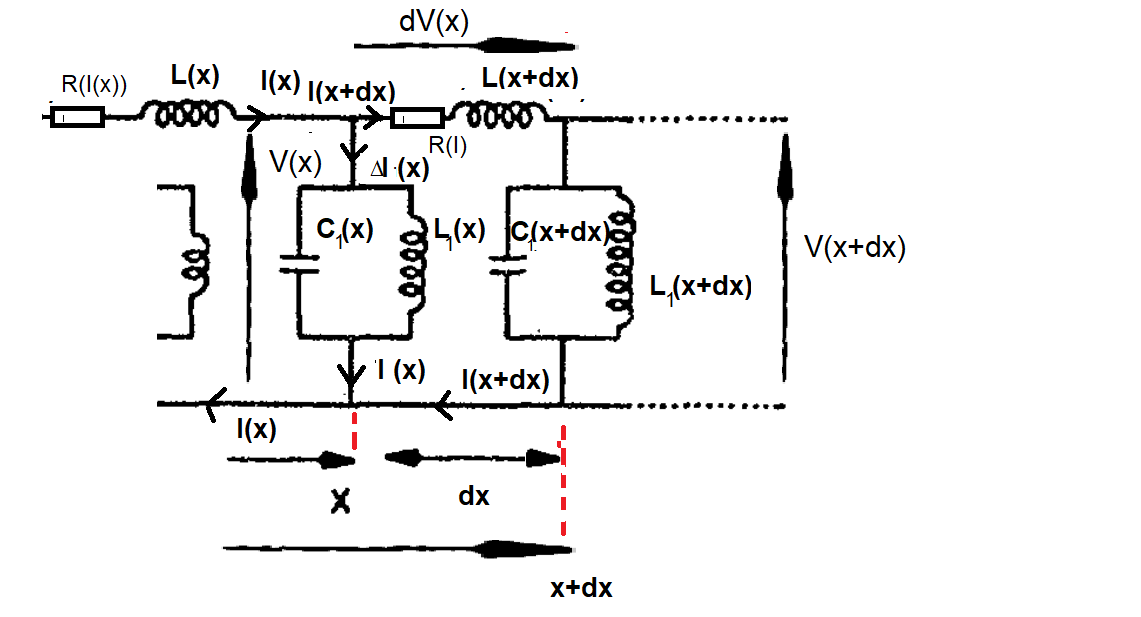}
\caption{Krons model in relation to long line model that is subjected to non-linear deformation introduced by non-linear resistive elements.}
\label{figs:NonLinearLongLine}
\end{figure}

We can deform the long line by introduction of non-linear resistive elements in series with inductance's places horizontally.
Let us refer to long-line model as depicted in Fig.\ref{figs:NonLinearLongLine} that can be characterized by  
equations
\begin{eqnarray}
dV(x)=dx \sqrt{-1}L\omega I(x), \nonumber \\
dI(x)=dx (\frac{1}{\sqrt{-1} L_1 \omega}+\sqrt{-1} \omega C_1) V(x)
\end{eqnarray}
which implies
\begin{eqnarray}
\frac{d}{dx}V(x)=\sqrt{-1} L \omega I(x), \nonumber \\
\frac{d}{dx}I(x)=(\frac{1}{\sqrt{-1} L_1 \omega}+\sqrt{-1} \omega C_1) V(x)
\end{eqnarray}
and that results in equation
\begin{eqnarray}
\frac{d^2}{dx^2}V(x)=\sqrt{-1}L \omega (\frac{1}{\sqrt{-1}L_1\omega}+\sqrt{-1}\omega C_1)V(x). \nonumber \\
\end{eqnarray}
Last equation can be rewritten to be of the form
\begin{eqnarray}
\frac{d^2}{dx^2}V(x)=(\frac{L}{L_1}-LC_1 \omega^2)V(x)=k_1(\omega)^2V(x), \nonumber \\
k_1(\omega)=\sqrt{(\frac{L}{L_1}-L C_1 \omega^2)}.
\end{eqnarray}
and has analytic solution as
\begin{eqnarray}
V(x)=a_1e^{+k_1x}+a_2e^{-k_1x}= \nonumber \\
=a_1e^{+\sqrt{(\frac{L}{L_1}-L C_1 \omega^2)}x}+a_2e^{-\sqrt{(\frac{L}{L_1}-L C_1 \omega^2)}x}
=b_1 sinh(k_1x)+b_2 cosh(k_1x)= \nonumber \\
=b_1 sinh(\sqrt{(\frac{L}{L_1}-L C_1 \omega^2)}x)+b_2 cosh(\sqrt{(\frac{L}{L_1}-L C_1 \omega^2)}x),\nonumber \\
I(x)=c_1e^{+k_1x}+c_2e^{-k_1x}= \nonumber \\
=c_1e^{+\sqrt{(\frac{L}{L_1}-L C_1 \omega^2)}x}+c_2e^{-\sqrt{(\frac{L}{L_1}-L C_1 \omega^2)}x}= \nonumber \\
=d_1 sinh(+\sqrt{(\frac{L}{L_1}-L C_1 \omega^2)}x)+d_2 cosh(-\sqrt{(\frac{L}{L_1}-L C_1 \omega^2)}x).
\end{eqnarray}

It is not hard to generalize obtained result for the case of R, L and C circuits components in series on upper cable branch and for the case of
circuits elements in parallel on inter-connecting branch with values of $R_1$, $L_1$ and $C_1$. 
In such a case we arrive to equations
\begin{eqnarray}
dV(x)=dx(\sqrt{-1}L\omega+R+\frac{1}{i\omega C})I(x), \nonumber \\
dI(x)=dx( \frac{1}{ \sqrt{-1} L_1 \omega } + \sqrt{-1}\omega C_1+\frac{1}{R_1}) V(x)
\end{eqnarray}
that can be decoupled, so one deals with two independent equations
\begin{eqnarray}
\frac{d^2}{dx^2}V(x)=(\sqrt{-1}L\omega+R+\frac{1}{i\omega C})(\frac{1}{\sqrt{-1}L_1\omega}+\sqrt{-1}\omega C_1+\frac{1}{R_1})V(x)=k_{1q}(\omega)^2V(x), \nonumber \\
\frac{d^2}{dx^2}I(x)=(\sqrt{-1}L\omega+R+\frac{1}{i\omega C})(\frac{1}{\sqrt{-1}L_1\omega}+\sqrt{-1}\omega C_1+\frac{1}{R_1})I(x)=k_{1q}(\omega)^2I(x).
\end{eqnarray}



In the next step we can incorporate nonlinear elements in previously considered long line models.
We have situation as depicted in Fig.\ref{figs:NonLinearLongLine}.
At first we consider linear circuit with linear density (per unit length) of $L(x)$, $L_1(x)$ and $C_1(x)$.
We have equations
\begin{eqnarray}
dV(x)=dx \sqrt{-1} L(x) \omega I(x), \nonumber \\
dI(x)=dx (\frac{1}{\sqrt{-1} L_1(x) \omega}+\sqrt{-1} \omega C_1(x))V(x)
\end{eqnarray}
what implies
\begin{eqnarray}
\frac{1}{\sqrt{-1} L(x) \omega }\frac{d}{dx}V(x)=I(x), \nonumber \\
\frac{d}{dx}(\frac{1}{\sqrt{-1}L(x)\omega }\frac{d}{dx}V(x))=(\frac{1}{\sqrt{-1}L_1(x)\omega}+\sqrt{-1}\omega C_1(x)) V(x).
\end{eqnarray}
Last equation implies
\begin{eqnarray}
(\frac{1}{\sqrt{-1}L(x)\omega})\frac{d^2}{dx^2}V(x)+(\frac{d}{dx}V(x))\frac{d}{dx}(\frac{1}{\sqrt{-1} L(x) \omega})=(\frac{1}{\sqrt{-1}L_1(x)\omega}+\sqrt{-1}\omega C_1(x))V(x).
\end{eqnarray}
and can be simplified to be
\begin{eqnarray}
(\frac{1}{L(x)})\frac{d^2}{dx^2}V(x)+(\frac{d}{dx}V(x))\frac{d}{dx}(\frac{1}{L(x)})=(\frac{1}{L_1(x)}-\omega^2C_1(x))V(x).
\end{eqnarray}
and we arrive to
\begin{eqnarray}
-(\frac{1}{L(x)})\frac{d^2}{dx^2}V(x)+(\frac{d}{dx}V(x))(\frac{d}{dx}L(x))(\frac{1}{L(x)^2})+\frac{1}{L_1(x)}V(x)=(+\omega^2C_1(x))V(x).
\end{eqnarray}
Under the assumption $L(x)=constants=L$ we have
\begin{eqnarray}
-(\frac{1}{L(x)})\frac{d^2}{dx^2}V(x)+\frac{1}{L_1(x)}V(x)=(+\omega^2C_1(x))V(x).
\end{eqnarray}
that has similarity with Schroedinger equation of the form
\begin{eqnarray}
-\frac{\hbar^2}{2m}\frac{d^2}{dx^2}\psi(x)+V_p(x)\psi(x)=E\psi(x).
\end{eqnarray}
indicating that energy E is related to $+\omega^2C_1(x)$ if we set capacitance that is independent of position, so $C_1(x)=const_2$.
Furthermore with presumption that V(x) is equivalent to wave-function $\psi(x)$ we can establish another analogies. We can spot that $\frac{1}{L_1(x)}$ plays of potential from Schroedinger equation $V_p(x)$, while $\frac{1}{L(x)}$ plays role of $\frac{\hbar^2}{2m}$.
Placement of linear resistance R in series with L inductance brings modification of previous equation into form
\begin{eqnarray}
-(\frac{1}{L(x)+\frac{R}{i \omega}})\frac{d^2}{dx^2}V(x)+\frac{1}{L_1(x)}V(x)=(+\omega^2C_1(x))V(x).
\end{eqnarray}
while placement of resistance $R_1$ in parallel to $L_1$ and $C_1$ brings  the equation
\begin{eqnarray}
-(\frac{1}{L(x)})\frac{d^2}{dx^2}V(x)+\frac{1}{L_1(x)+\frac{R1}{i \omega}}V(x)=(+\omega^2C_1(x))V(x).
\end{eqnarray}
and thus we have dissipative effective Schrodinger potential given as
\begin{eqnarray}
V_p(x)=\frac{1}{L_1(x)+\frac{R1}{i \omega}}=\frac{i \omega}{i \omega L_1(x)+R_1}=\frac{i \omega (-i\omega L_1(x)+R_1)}{ (\omega L_1(x))^2+R_1^2}=\frac{(\omega^2 L_1(x)+i R_1 \omega)}{ (\omega L_1(x))^2+R_1^2}.
\end{eqnarray}

\subsection{Case of non-linear long line model generalization of Kron's model}
We consider long line with case of vertical inductance dependence on space given by $L_1(x)$, while  preassuming other passive elements to be constant that leads to equation

\begin{eqnarray}
    \frac{d}{dx}I(x)=(j \omega C_1 + \frac{1}{j \omega L_1(x)})V(x), \nonumber \\
    \frac{d}{dx}V(x)=(j \omega L + I(x)R(I(x)))I(x),
\end{eqnarray}
and from previous form we obtain

\begin{eqnarray}
 \frac{d}{dx}I(x)=  \frac{d}{dx}(\frac{1}{(j \omega L + I(x)R(I(x)))} \frac{d}{dx}V(x))= (j \omega C_1 + \frac{1}{j \omega L_1(x)})V(x).
\end{eqnarray}
In general case we have
\begin{eqnarray}
    \frac{d}{dx}I(x,t)=(\frac{d}{dt}V(x,t)C_1- \int_{t_0}^{t} dt' \frac{1}{L_1(x)}V(x,t')), \nonumber \\
    \frac{d}{dx}V(x)=(-L\frac{dI}{dt} + I(x)R(I(x))),
\end{eqnarray}

In case of introduction of the non-linear resistance in series with upper inductors and under assumption of the uniformity of the long line
with explicitly given dependence of $L_1(x)$ we arrive into non-linear integro-differential equation.
Consequently we obtain
\begin{eqnarray}
    \frac{d^2}{dx^2}I(x,t)=(\frac{d}{dt}\frac{d}{dx}V(x,t)C_1- \int_{t_0}^{t} dt' \frac{d}{dx}[\frac{1}{L_1(x)}V(x,t')]), \nonumber \\
    \frac{d}{dx}V(x)=(-L\frac{dI}{dt} + I(x)R(I(x))),
\end{eqnarray}
that results in
\begin{eqnarray}
    \frac{d^2}{dx^2}I(x,t)=(\frac{d}{dt}C_1(-L\frac{dI}{dt} + I(x)R(I(x)))- \int_{t_0}^{t} dt' [\frac{d}{dx}\frac{1}{L_1(x)}]V(x,t'))- \int_{t_0}^{t} dt' \frac{1}{L_1(x)} [\frac{d}{dx}V(x,t')], \nonumber \\
    \frac{d}{dx}V(x)=(-L\frac{dI}{dt} + I(x)R(I(x))),
\end{eqnarray}
and applying operator $\frac{d}{dt}$ results in
\begin{eqnarray}
    \frac{d^3}{dx^2dt}I(x,t)=\frac{d}{dt}(\frac{d}{dt}C_1(-L\frac{dI}{dt} + I(x)R(I(x)))-  [\frac{d}{dx}\frac{1}{L_1(x)}]V(x,t'))-  \frac{1}{L_1(x)} [\frac{d}{dx}V(x,t')], \nonumber \\
    \frac{d}{dx}V(x)=(-L\frac{dI}{dt} + I(x)R(I(x))),
\end{eqnarray}
that has brings
\begin{eqnarray}
\frac{d}{dx} \Bigg[\frac{1}{[\frac{d}{dx}\frac{1}{L_1(x)}]}   [-\frac{d^3}{dx^2dt}I(x,t)+\frac{d}{dt}(\frac{d}{dt}C_1(-L\frac{dI}{dt} + I(x)R(I(x)))-  \frac{1}{L_1(x)} [-L\frac{dI}{dt} + I(x)R(I(x))]]\Bigg]=, \nonumber \\
    =(-L\frac{dI}{dt} + I(x)R(I(x)))=\frac{d}{dx}V(x),
\end{eqnarray}

and consequently we end up with non-linear partial differential equation for $I(x,t)$ of form

\begin{eqnarray}
 \frac{d}{dx}(\frac{1}{[\frac{d}{dx}\frac{1}{L_1(x)}]})]   [-\frac{d^3}{dx^2dt}I(x,t)+\frac{d}{dt}(\frac{d}{dt}C_1(-L\frac{dI(x,t)}{dt} + I(x,t)R(I(x,t)))-  \frac{1}{L_1(x)} [-L\frac{dI(x,t)}{dt} + I(x,t)R(I(x,t))]] \nonumber \\
 +\Bigg[ [(\frac{1}{[\frac{d}{dx}\frac{1}{L_1(x)}]})]   \frac{d}{dx}[-\frac{d^3}{dx^2dt}I(x,t)+\frac{d}{dt}(\frac{d}{dt}C_1(-L\frac{dI(x,t)}{dt} + I(x,t)R(I(x,t)))-  \frac{1}{L_1(x)} [-L\frac{dI}{dt} + I(x,t)R(I(x,t))]]\Bigg]  \nonumber \\
    =(-L\frac{dI(x,t)}{dt} + I(x)R(I(x,t)))=\frac{d}{dx}V(x), \nonumber \\
\end{eqnarray}

where L, R, $C_1$ are considered to be non-dependent on x and they represent upper inductance, linear or non-linear resistance and capacitance $C_1$, while $L_1(x)$ is inductance linear density that is explicitly position dependent.


In first approach it is instruction to consider non-linear resistance as represented by two identical diodes connected in anti-parallel way and in such a case the non-linear resistance is given by the formula:
\begin{eqnarray}
    R(I)=\frac{V}{I}=kT\frac{ln(\frac{I}{I_0})+1}{I}=\frac{V}{I_0 (e^{\frac{V}{kT}}-1)}=R(V).
\end{eqnarray}
Under the circumstance of low voltage $R(V \rightarrow 0)=\frac{kT}{I_0}$, while for non-small values of voltage we can assume $R(V)=\frac{V}{I_0} e^{-(\frac{V}{kT})}$.


\section{Conclusions}
$ $
\newline \newline
Conducted numerical simulations show potential of generalization of originally proposed by Kron's circuit towards representation of confined particle by different shapes of effective potential, like the harmonic, V-shape or polynomialy dependent.
\newline \newline
We believe that the creation of the mathematical model of the generalised solver based on Kron’s second circuit would enable us to create the dedicated simulation hardware. Such hardware could potentially vastly improve the speed and efficiency of the simulations of the quantum phenomena as one can conduct simulation with omission of existence of computers on local chips. Implementation of classical analog solvers in classical analog electronic circuitry can express certain class of quantum phenomena. Later this concept can be used in design of classical-quantum hardware.
One of the open issues is analog hardware simulation of two or more electrostatically interacting single electron semiconductor devices as given by \cite{DeformedQubit}. Another future targets can be focused on representing superconducting single-photon detectors \cite{TransitionEdge} and Josephson junction \cite{Washboard} by classical analog electronics.
\newline \newline
Based on already conducted simulations we can point out forthcoming research goals:
\begin{itemize}
    \item Creation of the fully generalised model based on Kron's second model of analog solver.
    \item Increasing the integration of the analog and digital electronics through use of the more advanced signal generators, voltage probes and data storage.
    \item Further automation of the simulation process.
\end{itemize}

\section{Acknowledgment}
This work was conducted equally in 50 percent by first Author and in 50 percent by second Author. The conceptual and mathematical scheme was drawn by second Author, while all conducted simulation were implemented and carried out by first Author.
\end{document}